# Observation of topologically distinct corner states in "bearded" photonic Kagome lattices

Limin Song[1†], Domenico Bongiovanni[1,2†], Zhichan Hu[1†], Ziteng Wang[1], Shiqi Xia[1], Liqin Tang[1], Daohong Song[1,3*], Roberto Morandotti[2], and Zhigang Chen[1,3**]

[1] *The MOE Key Laboratory of Weak-Light Nonlinear Photonics, TEDA Applied Physics Institute and School of Physics, Nankai University, Tianjin 300457, China*

[2] *INRS-EMT, 1650 Blvd. Lionel-Boulet, Varennes, Quebec J3X 1S2, Canada*

[3] *Collaborative Innovation Center of Extreme Optics, Shanxi University, Taiyuan, Shanxi 030006, China*

[†]*These authors contributed equally to this work*

*\*songdaohong@nankai.edu.cn; \*\*zgchen@nankai.edu.cn*

**Kagome lattices represent an archetype of intriguing physics, attracting a great deal of interest in different branches of natural sciences, recently in the context of topological crystalline insulators. Here, we demonstrate two distinct classes of corner states in breathing Kagome lattices (BKLs) with "bearded" edge truncation, unveiling their topological origin. The in-phase corner states are found to exist only in the topologically nontrivial regime, characterized by a nonzero bulk polarization. In contrast, the out-of-phase corner states appear in both topologically trivial and nontrivial regimes, either as bound states in the continuum or as in-gap states depending on the lattice dimerization conditions. Furthermore, the out-of-phase corner states are highly localized, akin to flat-band compact localized states, and they manifest both real- and momentum-space topology. Experimentally, we observe both types of corner states in laser-written photonic bearded-edge BKLs, corroborated by numerical simulations. Our results not only deepen the current understanding of topological corner modes in BKLs, but also provide new insight into their physical origins, which may be applied to other topological BKL platforms beyond optics.**

Keywords: topological corner states, bearded Kagome lattices, momentum-space topology, real-space topology, photonic lattice

Periodic lattices with a particular geometry and topology are of great interest in the study of fundamental phenomena in different branches of physics. A typical example is the Kagome lattice [1], which has turned into one of the most studied models in condensed matter physics, ultracold atoms, as well as photonics [2-15]. For instance, tremendous research efforts involving Kagome lattices have led to the understanding of a host of intriguing phenomena, ranging from the properties of fractional quantum Hall states [3], quantum spin liquids [4, 16], charge order and superconductivity [9], to compact localized states [6], anomalous Landau levels [7, 15], and flat-band exciton-polariton emission [10]. As a typical example in photonics, a Corbino-shaped Kagome lattice has been realized [12] for direct observation of flatband non-contractible loop states (NLSs) protected by real-space topology, originally predicted from the "frustrated" hopping models [2].

Recently, the so-called breathing Kagome lattices (BKLs) [17] have been employed as a prototypical platform to study higher-order topological insulators (HOTIs) [18, 19], realized in a variety of systems [20-23]. The HOTI features germinating from the BKLs have also been extended to the nonlinear regime [22], as well as to the square-root HOTIs for which the Kagome lattice acts as a "parent" lattice [24-26]. Quite recently, $p$-orbital HOTIs based on the BKLs have been proposed and demonstrated in photonics [27, 28].

Whilst the number of experimental realizations of HOTI corner states based on BKLs is still increasing, the origin and classification with respect to their higher-order topology have been extensively investigated. It has been established that the quantized bulk polarization typically acts as an effective topological invariant for the HOTIs [17, 18, 29]. In a two-dimensional (2D) $C_n$-symmetric topological crystalline insulator, a zero-dimensional corner state is often described as of "higher-order" topology following the convention that the codimension is two, but later theoretical analysis and experimental observations have shown that a fractional corner anomaly is the key to reveal higher-order topology [19]: a topological crystalline insulator can be classified as a HOTI when the fractional corner anomaly is nonzero, while it is not a HOTI when such a quantity is zero. Nevertheless, there are still ongoing debates about the HOTI nature of the corner states in BKLs [30-32]. Recently, topological protection of corner states in BKLs has been studied by using the concept of sub-symmetry and long-range hopping symmetry [33].

In a BKL, there are two different choices of unit cells, upward- and downward-pointing triangles, which are derived from two distinct types of primitive generators for the $C_3$-symmetric topological crystalline insulator [29]. A downward-pointing unit cell is illustrated in Fig. 1(a). According to the boundary configurations of the resulting lattice, a finite-sized BKL can be cut with "flat" [19-22, 33-36] or "bearded" edges [see Fig. 1(a)], depending on the appropriate choice of unit cells. Thus far, much of the previous exploitations of HOTI corner states in BKLs have focused on its flat-edge structures, while little attention has been devoted to the bearded-edge ones. Only a few theoretical works in photonic crystals reported on the existence of corner states in the bearded-edge BKLs [23, 37]. Nevertheless, the underlying physical origin of such corner states is either inappropriately classified as trivial topology [37] — mainly resulting from the confusion of the bulk-boundary correspondence, or simply interpreted as HOTI states [23] — a conclusion mostly based on the nonzero value of bulk polarization. A clear understanding of the topological origin of these corner states is still lacking, and their experimental observation in bearded BKLs has not been realized whatsoever.

In this work, we investigate theoretically and demonstrate experimentally two distinct classes of corner states in the bearded BKLs: "in-phase" and "out-of-phase" (or symmetric and antisymmetric) corner states. From the intrinsic characteristics of the BKL as well as the intuitive picture that the corners are "boundaries of boundaries" [18], we find that these two types of corner states arise differently and enjoy different topological protections. Specifically, the in-phase corner states (IPCSs) exist only in the nontrivial regime and are protected by momentum-space topology, characterized by a nonzero bulk polarization. In contrast, the out-of-phase corner states (OPCSs) appear in any dimerization regime of a BKL structure and are protected by both momentum- and real-space topology. Associated with strongly localized flat-band states in both trivial and nontrivial regimes, their corresponding eigenvalues reside either in the bulk continuum, suggesting the formation of bound states in the continuum [38], or in the gap. Both of these classes of corner states are observed directly in photonic BKLs with bearded edges established with a CW-laser-writing technique, and the results are further corroborated by numerical simulations.

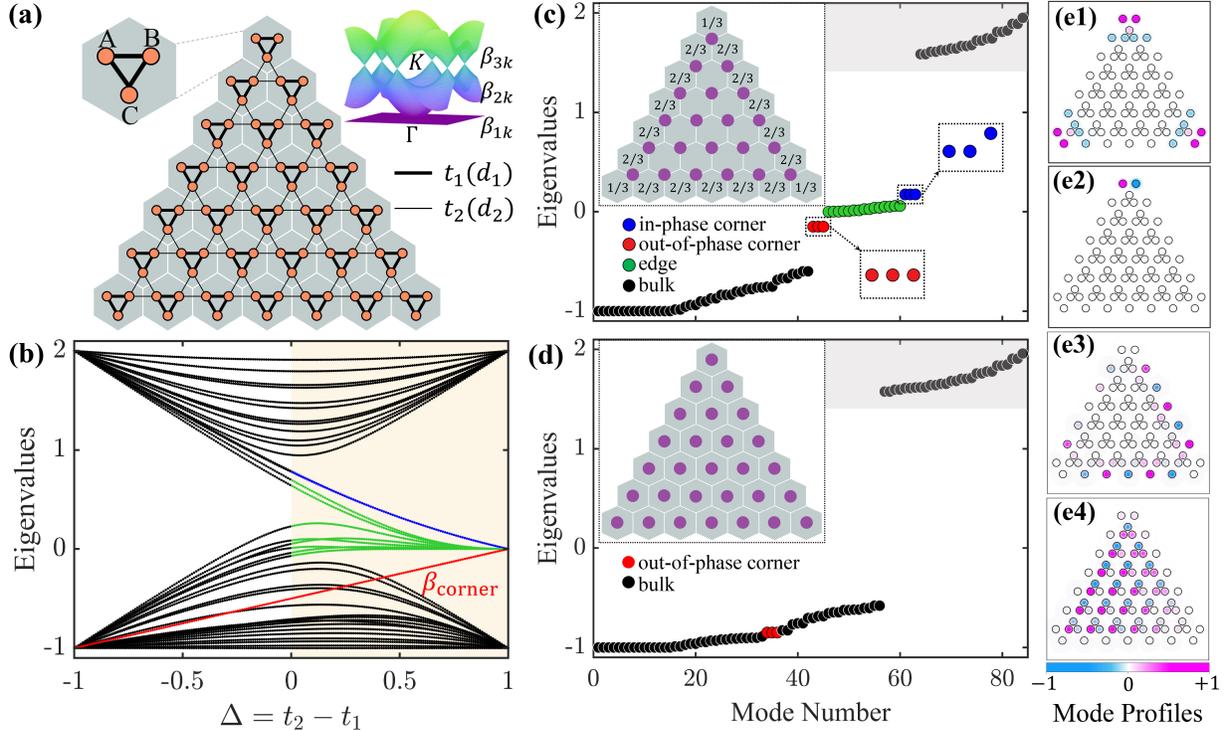

**Fig. 1: Existence of two types of corner states in photonic "bearded" BKLs.** (a) Illustration of a finite "bearded" BKL with 28-unit cells (shaded hexagons), where $t_1$ and $t_2$ are the intra- and inter-cell coupling strengths, tuned by lattice spacings $d_1$ and $d_2$, respectively. Left inset marks one unit cell composed of three sublattice sites (A, B, and C), while right inset illustrates the band structure of its corresponding infinite lattice with K and Γ being the high-symmetry points. (b) Eigenvalue spectrum plotted as a function of the dimerization parameter $\Delta = t_2 - t_1$, highlighting topologically trivial ($\Delta<0$, blank background) and nontrivial ($\Delta>0$, colored background) regimes, as identified by the bulk polarization for the BKL. (c) Calculated band diagram at $\Delta=0.7$, where the left inset illustrates the fractional mode density (mod 1) of each unit cell for the shaded band. Purple circles denote the Wannier centers. Zoom-in plots of out-of-phase and in-phase corner-state eigenvalues shown in dashed squares reveal their "perfectly" and "nearly" degenerate features, respectively. (d) Same as (c), for the trivial regime at $\Delta = -0.7$. (e1-e4) Representative mode profiles from (c), showing (e1) in-phase and (e2) out-of-phase corner states, as well as (e3) edge and (e4) bulk states.

Firstly, we focus on the band structure and eigenvalue calculations of a bearded BKL with downward triangles as unit cells (see Fig. 1(a)). Under the tight-binding approximation, light propagation through the lattice (consisting of evanescently coupled waveguides) along the $z$-direction is given by

$$i\partial_z |\psi_n(z)\rangle = H_{\text{TB}}(t_1, t_2)|\psi_n(z)\rangle, \qquad (1)$$

where $|\psi_n(z)\rangle$ is the light amplitude at the $n$-th site, and $H_{\text{TB}}$ is the Bloch Hamiltonian that depends on the inter- and intra-cell coupling coefficients $t_1$ and $t_2$ — see definition in Supplementary Information (SI). The dimerization parameter $\Delta = t_2 - t_1$ is employed to "define" the nontrivial and trivial geometries, as commonly used for the BKLs [22, 34, 36]. In photonic structures as established in our experiment, the couplings $t_1$ and $t_2$ can be flexibly tuned by changing the relative spacing $d_1$ and $d_2$ between the corresponding waveguides, respectively. We first numerically calculate the eigenvalues at various dimerization conditions and plot them in Fig. 1(b), along with two representative spectra for the nontrivial [Fig. 1(c)] and trivial [Fig. 1(d)] geometries. The topologically nontrivial ($\Delta > 0$) and trivial ($\Delta < 0$) BKLs are characterized by different values of bulk polarization $\boldsymbol{P}$ = (-1/3, -1/3) and $\boldsymbol{P}$ = (0, 0), respectively (see SI for details).

Different from flat BKL models [17, 19-21, 34-36], the bearded BKLs can simultaneously host two distinct types of corner modes in the nontrivial regime [Fig. 1(c)]: namely IPCSs and OPCSs (blue and red circles), which have nonzero eigenvalues and are separated from each other by an edge band (green circles). To highlight the difference between the two types of corner modes, their dimensionless amplitude and phase profiles in the nontrivial regime with $\Delta = 0.7$ are illustrated in Figs. 1(e1, e2) and compared with those of edge and bulk states in Figs. 1(e3, e4). One can see that the IPCS populates many lattice sites around the three corner regions and decreases exponentially with a staggered phase when going to the bulk [Fig. 1(e1)]. In contrast, the OPCS only confines to the outermost two sites of the lattice vertex with an opposite phase relation [Fig. 1(e2)]. Corner states in HOTIs reported in previous works [20-22, 39] are usually sensitive to the lattice size and present a peculiar mode distribution, occupying specific sublattices with an appropriate phase relation among neighbor cells. In the bearded BKL, however, the properties manifested by the corner modes are rather distinctive from those in HOTIs and merit further theoretical exploration. In this work, we focus mainly on the experimental realization, along with an intuitive explanation.

Next, we present experimental results for the observation of predicated topological corner states in photonic bearded BKLs, established in a 20 mm-long biased photorefractive crystal (SBN:61) via the laser writing technique [40]. Since the IPCSs and OPCSs are three-fold degenerate due to the intrinsic $C_3$-rotational symmetry exhibited by this triangular Kagome structure, the corner states can appear at any lattice vertex, or distribute to three physically equivalent corners. In the experiment, without loss of generality, the corner states are excited by employing a dipole-like probe beam that matches the mode distribution at the uppermost sites of the top lattice vertex (see white arrows in Fig. 2(a1)). We structure the input dipole-like probe beam for the following reasons: i) For the in-phase corner modes, the intensity ratio between the C site at the top corner (or other bulk sites) and the dominant A and B sites is less than 2.56% in our structure (Fig. 1(e)), small enough so not to change the underlying physics; ii) For the out-of-corner counterparts, the probe beam occupies only the two uppermost sites [Fig. 1(e2)] — thereby a direct experimental comparison between the two distinct types of corner states can be presented. Results for the observation of IPCSs are shown in Fig. 2.

As illustrated in Fig. 2(a2), the output intensity pattern of the in-phase probe beam perfectly localizes at the initially excited corner sites A and B for the nontrivial lattice, without any significant coupling to the nearest-neighbor site C. Instead, an evident coupling of light to the third waveguide C is observed for both uniform and trivial lattices under in-phase excitation [Figs. 2(b2) and 2(c2)]. Our experimental results are in good agreement with corresponding numerical simulations, as shown in Figs. 2(a3-c3), confirming the existence of IPCSs only in the topological nontrivial regime. Long-distance evolutions of the in-phase probe beam with 20% amplitude perturbation at one of the top two corner sites for the three lattice configurations are shown in Figs. 2(a4-c4). Specifically, in the nontrivial lattice, a small portion of energy couples to the third waveguide C of the top corner unit cell for long-distance propagation, but most of the energy concentrates at the two uppermost sites A and B [Fig. 2(a4)]. That happens because the energy in the eigenmode distribution of the IPCSs is mostly confined to the top two sites while having exponential tails in the bulk [Fig. 1(e1)], and we only excite the two uppermost waveguides that do not match the corner mode perfectly. In the uniform and trivial lattices, such an in-phase probe beam experiences discrete diffraction, with most of the light spreading into the bulk [Figs. 2(b4) and 2(c4)], again suggesting that IPCSs cannot exist in trivial geometries.

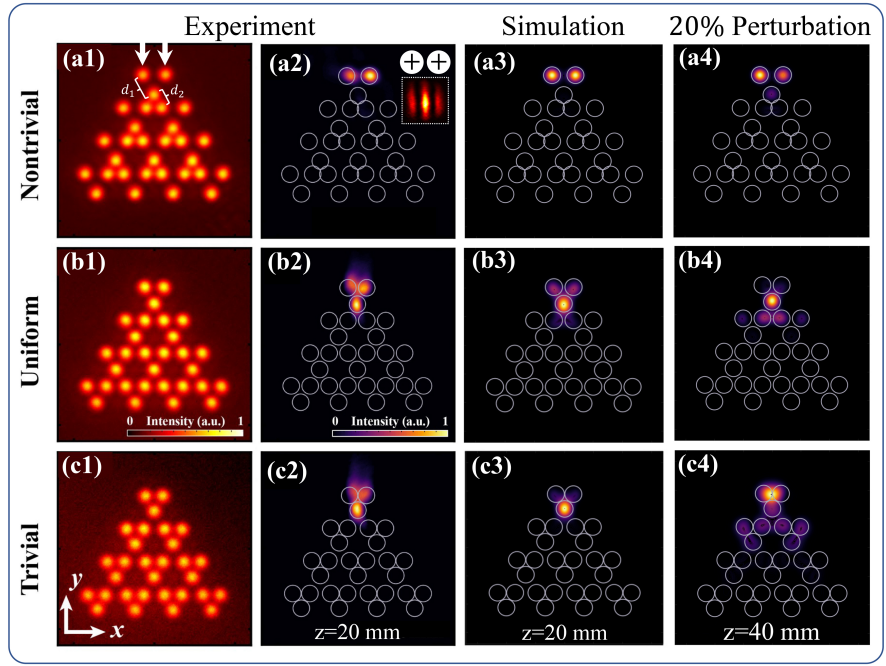

**Fig. 2: Observation of in-phase corner states in photonic "bearded" BKLs.** (a1-c1) Experimentally established (a1) topologically nontrivial, (b1) uniform, and (c1) topologically trivial lattice structures, obtained by a site-to-site laser-writing technique in a nonlinear crystal. White arrows in (a1) mark the excited waveguides at the input facet. The interference pattern of the dipole-like probe beam shown in the dashed square of (a2) indicates the in-phase relation, manifested by the bright fringes in the center due to constructive interference. (a2-c2) Experimental output intensity patterns of in-phase probe beams for (a1) topological, (b1) uniform, and (c1) trivial

lattice geometry. (a3-c3, a4-c4) Numerical results of output intensity patterns obtained for different propagation distances: $z = 20$ mm (a3-c3) and 40 mm (a4-c4). The latter cases with 20% amplitude perturbation at one of the top two corner sites. Experimental and numerical parameters are: $d_1 = 40$ μm, $d_2 = 27$ μm for (a1-a4); $d_1 = d_2 = 33.5$ μm for (b1-b4); and $d_1 = 30$ μm, $d_2 = 37$ μm for (c1-c4). Due to the two-site excitation and limitations in the crystal length (20 mm), the fine feature of the corner mode as displayed in Fig. 1(e1) is not directly observed in the experiment but numerically verified by performing long-distance continuous-model simulations. Another set of experiments with a six-site excitation that covers three uppermost unit cells, thus matching better the mode profile in Fig. 1(e1), is also provided in Supplementary Information.

We perform another set of experiments with an out-of-phase dipole-like corner excitation to observe the OPCSs. Since their mode profiles are completely localized at the corner-most sites of the lattice and have strictly zero amplitude elsewhere [Fig. 1(e2)], they are perfect "compact corner states" [11, 41, 42]. However, such compact corner states are essentially different from all previously reported topological corner states that typically exhibit "exponential-decay" details, for which we shall discuss the underlying physical picture later. To demonstrate this distinctive compact feature of OPCSs, the probe beam is modulated to match the corner mode and is launched to excite the two uppermost sites of the lattice. In the topologically nontrivial geometry, the output intensity pattern is localized at the initially excited corner sites A and B without coupling to the nearest-neighbor site C, as shown in Fig. 3(a1). Numerical simulations carried out by using the same experimental parameters are in good agreement with observations [Fig. 3(a2)]. Due to the limited length (20 mm) of the crystal, the propagation dynamics of the OPCSs for a long distance are also investigated numerically, as shown in Fig. 3(a3). One can see that even in the condition of 20% amplitude perturbation, the corner state remains robust during propagation. Moreover, we also experimentally explored the aptitude of OPCSs to survive in other lattice configurations. One can see that even for the uniform and the trivial geometries, OPCSs remain strongly localized at the initially excited A and B waveguides, demonstrating their intrinsic resilience in excellent agreement with both the numerical [Figs. 3(b2, b3) and 3(c2, c3)] and theoretical results shown in Fig. 1.

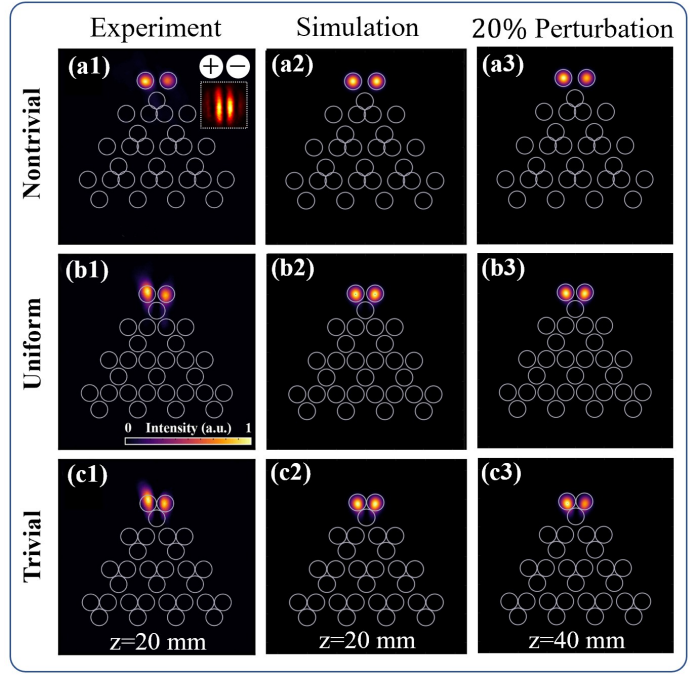

**Fig. 3: Observation of out-of-phase corner states in photonic "bearded" BKLs.** The lattices for the three regimes are the same as shown in the first column of Fig. 2. (a1-c1, a2-c2) Output intensity patterns of out-of-phase probe beam after 20 mm-long propagation obtained from (a1-c1) experiments and (a2-c2) numerical simulations. The interference pattern with a central dark fringe shown in the dashed square of (a1) indicates the out-of-phase relation of the dipole-like probe beam. (a3-c3) Output intensity patterns for a propagation distance of $z = 40$ mm, but with 20% amplitude perturbation at one of the top two corner sites. In all cases, the out-of-phase corner states remain intact.

In what follows, we provide a detailed theoretical analysis of the topological properties and the underlying physical mechanism of two topologically distinct corner states observed in bearded BKLs. The IPCSs are strictly connected to the presence of a topologically nontrivial phase (see blue line in Fig. 1(b)) and exhibit nearly degenerate eigenvalues. Since they share their birth (death) with the presence (absence) of a nontrivial regime divided by bulk polarization, the topological properties of IPCSs originate from the momentum space. This character is closely related to the gap closing and reopening occurring at the six Dirac points of two uppermost dispersive bands [Fig. 1(a)] (see SI). Therefore, IPCSs are connected to the appearance of the topologically nontrivial edge states localized in the bandgap [17, 43]. However, although the bulk polarization $P$ is nonzero for $\Delta > 0$, they are not HOTI states since the fractional corner anomaly $\phi$ is null in this case [19]. The latter is defined as $\phi = \rho - \sigma_1 - \sigma_2$ (mod 1), where $\rho$ is the fractional mode density localized to the corner unit cell, while $\sigma_1$ and $\sigma_2$ are those manifested at the edge unit cells that intersect to form the corner [19]. Calculations of fractional mode density for nontrivial bearded BKL ($\Delta = 0.7$) are shown in the inset of Fig. 1(c) where one can find that $\phi \approx 1/3 - 2/3 - 2/3 = 0$ (mod 1). A zero value of fractional corner anomaly indicates that the corner states are not of higher-order topological origin [19]. Thus, the topological origin of IPCSs is different from those demonstrated in 2D SSH lattices [39, 44] and flat-edge BKLs [20-22, 34], which have both nonzero $\phi$ and zero-energy topological states [19, 29].

On the other hand, OPCSs have perfectly degenerate eigenvalues $\beta_{\text{corner}}$ (see red line in Fig. 1(b)) and exist in both topologically trivial and nontrivial regimes, by appearing firstly in the bulk continuum and then in the gap for highly nontrivial regimes ($\Delta > 0.5$). In particular, the corner mode density and fractional corner anomaly of the OPCSs are zero in the trivial regime [Fig. 1(d)], suggesting a different topological origin for these corner states. Interestingly, we note that the existence of the OPCSs in any topologically trivial and nontrivial regimes resembles the comportment of the lowest two energy bands of the BKL model that experiences no close-and-reopen transition during the dimerization process. Consequently, their physical origin and topological properties must be comprehended from the different perspectives of momentum- and real-space topologies. Intuitively, the characteristic features of OPCSs can be ascribed to an inheritance from the noncontractible loop states (NLSs) [Fig. 4(a)], arising from singular flat-band touching at the $\Gamma$ point [inset in Fig. 1(a)] [2, 12, 45, 46]. In this regard, since the bearded-edge truncation procedure of the infinite BKL structure enables the intersections of three broken NLSs at the corners, as sketched in the hexagon-tiling region of Fig. 4(a), it is straightforward to assume that the topological nature of OPCSs is related to NLSs (or flat band with singular band-touching). A quantitative description of this assertion can be provided by carefully looking at the linear evolution of fully degenerate propagation constants (i.e., eigenvalues) associated with the three OPCSs in Fig. 1(b), reading as $\beta_{\text{corner}} = (\Delta - 1)/2 = -t_1$. In the momentum space, the flat-band $\beta_{1k}$ is located at $-t_1 -$ $t_2$, which is coincident with the propagation constant of the NLSs along three different directions in Fig. 4(a). The bearded-edge truncation of the BKL suddenly eliminates the effect of the inter-cell coupling $t_2$ outside of the hexagon-tiling triangular region, thereby leaving three OPCSs [Fig. 4(a)]. In essence, the relation

$$\beta_{\text{corner}} = \beta_{1k} + t_2 \tag{2}$$

indicates the intimate connection between them and directly manifests the flat-band-related origin of the OPCSs.

Now, the connection between OPCSs and singular band touching is examined by the outcomes of flat-band perturbations, which confirm the need for additional real-space topological protection to preserve their special features, including compact localization in only two sites and perfect eigenvalue degeneracy. To this end, the following quantity is introduced

$$\eta_s = \left|\left\langle \psi^{(s)}_{\text{corner}} \middle| \psi^{(\delta)}_{\text{corner}} \right\rangle\right|^2, \tag{3}$$

where the parameter $\eta_s$ ($s$ denotes A-B, B-C, or A-C corners) ranges between 0 and 1 and returns the degree of similarity between the corner mode distributions under perturbed $\left|\psi^{(\delta)}_{\text{corner}}\right\rangle$ and unperturbed conditions $\left|\psi^{(s)}_{\text{corner}}\right\rangle$. In addition, the energy gap between the lower two bands is defined as $\Delta\beta_k = \min(\beta_{2k} - \beta_{1k})$, where $\beta_{1k}$ is the flat-band energy (or the uppermost energy value for the perturbed flat band) and $\beta_{2k}$ is the lowermost energy level of the middle dispersion band (see inset in Fig. 1(a)). The flat-band touching point $\Gamma$ is protected by real-space topology and can only be removed by perturbations that destroy the flat band [2, 45]. One way to open a band gap between the lower two bands is to apply on-site perturbations. When the flatness of the low-energy band is destroyed, the OPCSs are expected to completely disappear or at least exhibit a certain instability. Here, we investigate perturbations affecting the flat-band touching point but preserving the local symmetry in the B-C out-of-phase corner mode. This is achieved by increasing the perturbation strengths $\delta_A$, which is only applied to the A sublattice, and maintaining an equal on-site potential $\delta_B = \delta_C = 0.15$ at the B and C sites. The numerical evolutions of $\eta_s$ and $\Delta\beta_k$ are shown in Fig. 4(b). For $\delta_A = 0.15$, the $C_3$ and global mirror symmetries of the system are preserved, and both flat band and singular band touching are present ($\Delta\beta_k = 0$, green arrow in Fig. 4(b)). This is equivalent to shifting the corner states by a constant energy, which does not change the underlying physics and is confirmed by all the preserving corner states, having a high degree of similarity with corresponding unperturbed ones ($\eta_{\text{all}} = 1$, red arrow in Fig. 4(b)). If $\delta_A \neq 0.15$, the flat-band touching at the $\Gamma$ point splits due to the applied perturbation, and a $\delta_A$-depending bandgap occurs ($\Delta\beta_k \neq 0$ in Fig. 4(b)) when compared to the unperturbed BKL. As expected, $\eta_s$ reaches its maximum value exactly where $\Delta\beta_k(\Gamma)$ becomes zero (dashed line in Fig. 4(b)). Note that preserving the global mirror symmetry is not a necessary condition for the existence of OPCSs because they can also exist when the global mirror symmetry is broken. If the singular flat-band touching point vanishes, the triply degenerate OPCSs disappear and spread into the bulk for larger perturbations (captured by $\eta_{AC} < 1$ and $\eta_{AB} < 1$ in Fig. 4(b)), unless an equal on-site potential is provided at the corner-most sites (B and C here) so as to preserve local symmetry ($\eta_{BC} \equiv 1$ in Fig. 4(b)). These numerical results not only reveal the underlying connection between the OPCSs and the singular flat-band touching point but also suggest that the out-of-phase compact corner modes are protected by local mirror symmetry. More details for the aforementioned symmetries can be found in the SI.

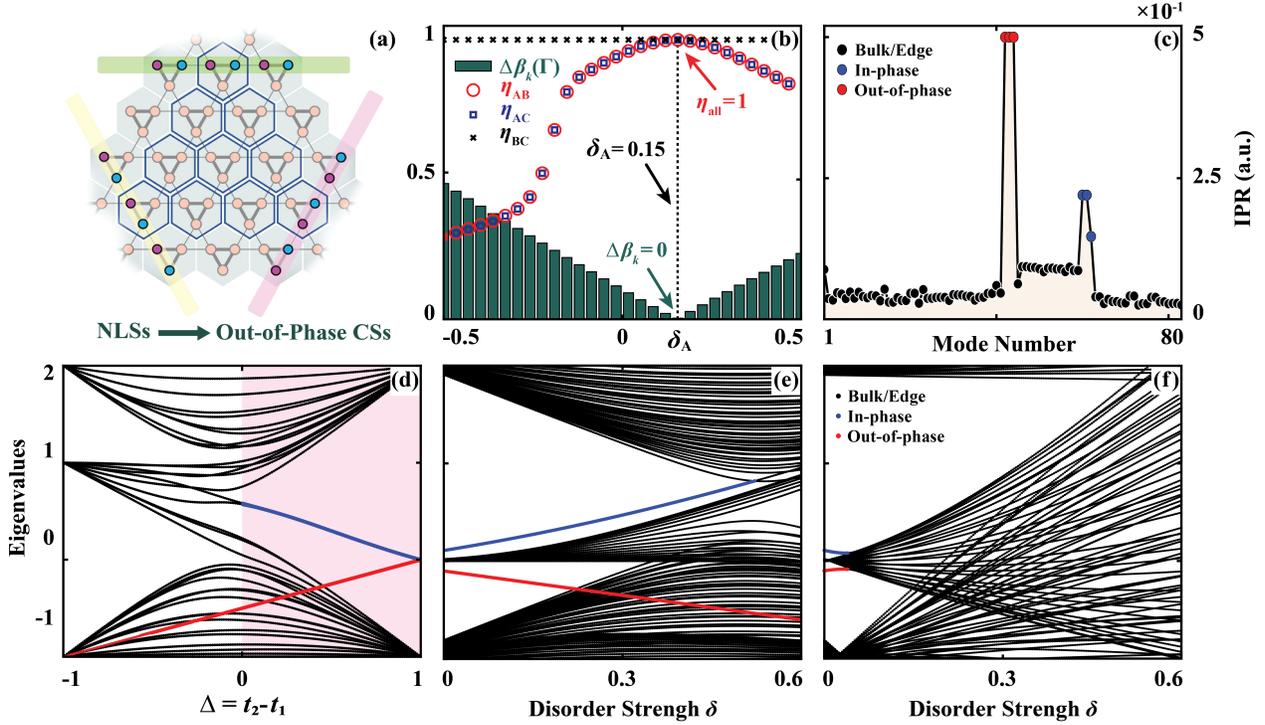

**Fig. 4: Theoretical analysis and perturbation test for the "bearded" BKL corner states.** (a) Schematic diagram to illustrate the formation of out-of-phase corner states from the NLSs. Magenta (peacock blue) circles indicate lattice sites with nonzero amplitude +1 (-1), while orange circles are the ones with vanishing amplitude. (b) Connection between the existence of out-of-phase corner states and flat-band touching. The left axis is the bar chart for the parameter $\Delta\beta_k$, the energy gap between the lower two bands $\beta_{1k}$ and $\beta_{2k}$ at the $\Gamma$ point in Fig. 1(a). The right axis is for the parameter $\eta_s$, the projection function of perturbed to unperturbed corner modes B-C, A-C, and A-B, as defined in the text, where $\eta_{BC}$, $\eta_{AC}$, and $\eta_{AB}$ are distinguished by cross, square, and circle, respectively. These two parameters are plotted as a function of the on-site perturbation $\delta_A$ for all A sublattices. $\delta_B = \delta_C = 0.15$. (c) IPR plot of modes in nontrivial regime with $\Delta = 0.7$. (d) Eigenvalue spectrum — similar to that in Fig. 1(b) but for the lattice structure *without* bearded-edge sites. (e) and (f) Robustness and stability analysis of the two types of corner states under perturbation tests preserving (e) the generalized chiral symmetry, $C_3$-rotational symmetry, and mirror symmetry, and (f) the $C_3$ and mirror symmetries but not the generalized chiral symmetry.

The perfect degeneracy of OPCSs' eigenvalues is another characteristic that distinguishes them from their in-phase counterparts. Indeed, it is a clear signature of the exclusive mode localization at the two vertex sites [Fig. 1(e2)], revealing a certain peculiarity deriving from a local destructive interference that effectively decouples them from the rest of the bulk lattice. This behavior is analogous to that of the antisymmetric bound states in the continuum, which are embedded in and decoupled from the continuum bulk states [47]. The localization properties of corner states can be quantified by evaluating the standard inverse participation ratio [48], defined as IPR= $\sum |\psi_n|^4 /(\sum |\psi_n|^2)^2$. Highest peaks in the IPR curve indicate the existence and degree of confinement of the corner modes. As shown in Fig. 4(c), the IPR curve obtained from the nontrivial bearded BKL reveals a much higher degree of localization for the OPCSs (red circles) as compared to the IPCSs (blue circles). Remarkably, an IPR value of 0.5 not only indicates that their mode distributions equally populate only two lattice sites [49] in the nontrivial BKL, but it also attests that OPCSs are truly localized compact states.

The difference between the two types of corner states and their topological protection is more transparent when involving the effect of removing bearded-edge sites. As shown in Fig. 4(d), the curve describing the eigenvalue evolution of the IPCSs (blue line) in the nontrivial regime appears more linear when compared with the same in Fig. 1(b), in the presence of the bearded-edge band, while the OPCSs (red line) remain invariant. This suggests that bearded edges indeed affect the IPCSs, but they have nothing to do with the OPCSs, further supporting that OPCSs have different origins. We also emphasize that there are differences between the OPCSs and the conventioanl CLSs: the former can compactly localize at the corners with nontrivial topology, whereas the latter can localize anywhere in the lattice with their eigenvalues always forming a flatband but no topological feature in momentum space.

One of the most important effects of topological protection on corner modes is their robustness against perturbations preserving certain types of symmetry [50, 51]. In Figs. 4(e) and 4(f), we numerically plot the results of robustness and stability analysis for the two types of corner states in the nontrivial regime, performed by introducing random perturbations to the $H$'s hopping ($H$ is the Hamiltonian for the finite bearded BKL, see SI). The applied perturbation matrix can be simply expressed as

$$H_d = \sum_{\langle i,j \rangle} \delta\, d_{ij} c_i^\dagger c_j, \tag{4}$$

where $i$ and $j$ ($i \neq j$) are the site indexes, $\delta$ is the perturbation strength, $d_{ij}$ is randomly distributed between $-1$ and 1. Corner-state robustness is examined by the effect of increasing the magnitude of $\delta$ on their eigenvalue evolution. Numerical results show that for perturbations maintaining the generalized chiral, $C_3$, and mirror symmetries, IPCSs preserve well in the bandgap until it closes [Fig. 4(e)]. Instead, the OPCSs merge into the bulk continuum before the gap closes and form BICs. For perturbations maintaining the $C_3$ and mirror symmetries but not the generalized chiral symmetry (broken by next-nearest-neighbor couplings between A-A, B-B, and C-C sites [33]), both types of the corner states vanish before the gap closing takes place [Fig. 4(f)]. A direct confirmation of this robustness test is provided by calculating the IPR curves of the perturbed bearded BKL for any $\delta$ values, as shown in SI. Based on our above analyses, IPCSs and OPCSs share in common momentum-space topological protection from $C_3$-rotational and generalized chiral symmetries, exhibiting nonzero-energy mode features. Additionally, OPCSs require protection from real-space topology to preserve distinctive characteristics of perfectly eigenvalue degeneracy, with compact amplitude localization at two corner sites only. These results agree well with our theoretical analyses and experimental observations.

In conclusion, we have theoretically and experimentally demonstrated nontrivial corner states in photonic BKLs with bearded-edge truncation, unveiling two classes of corner states that feature distinct phase structures and topological origins: the IPCSs from exclusively momentum-space topology and the OPCSs from both momentum- and real-space topology. An intriguing feature of the latter is that they can convert from BICs to in-gap states. This characteristic is in sharp contrast to the HOTI corner modes, and the physical origin can be found in the highly localized states protected by local spatial symmetry and topology [47, 52]. Our results not only provide a versatile platform to study topological corner states but also bring about new opportunities to explore fundamental physics arising from the interplay between momentum-space and real-space topology. Moreover, with growing interest in various realistic materials with (breathing) Kagome arrangements [20-22, 33-36], topologically distinct corner states presented here may be applicable to other systems. Our work may also prove relevant to future photonic device applications including for example topological corner-state lasing [53] and error filtering for quantum qubits [54].

## Acknowledgments

We thank Hrvoje Buljan for the helpful discussions. This research is supported by the National Key R&D Program of China (No. 2022YFA1404800); the National Natural Science Foundation of China (Nos. 12134006, 11922408, 12374309, and 12274242); the Natural Science Foundation of Tianjin (No. 21JCYBJC00060 and No. 21JCJQJC00050); and the 111 Project (No. B23045) in China. D.B. acknowledges support from the 66 Postdoctoral Science Grant of China and the National Natural Science Foundation (12250410236). R.M. acknowledges support from NSERC and the CRC program in Canada.